**Transformation fields in the Extended Space Model – prediction and secondary test experiment.**

Tsipenyuk D.Yu.
General Physics Institute of the Russian Academy of Sciences
Vavilova str.38, 119991, Moscow, Russia.
E-mail: tsip@kapella.gpi.ru

*Repeated series of experiments by checking of a prediction concerning capability of gravitational field generation at braking of charged massive particles in the substance are conducted. The prediction was made within the framework of the extended 5-D space model with the metric (1 + 4).*
*The new scheme of experiment has allowed receiving results that qualitatively coincide with results of the first experimental series.*
*Quantitatively in comparison with the first series in repeated experimental series results increased more than in 60 times. Additional force equal to $6,7*10^{-5}$ N was reached in experiments with stopping charged massive particles in the substance.*

## Introduction

In [1,2] was advanced 5-D extended space model (ESM). ESM generalizes a special theory of relativity on 5-D extended space with the metric (+; -, -, -, -). Model integrating electromagnetic and gravitational interaction was constructed in these papers.

For this purpose extension of (1 + 3)-D Minkowski space M (t; x, y, z) on (1 + 4) -D extended space was constructed.

As 5-th additional coordinate interval s is used. Interval s already exists in the Minkowski space. The coordinates in (1 + 4) -D extended space are connected by well-known ratio:

$$s^2 = (ct)^2 - x^2 - y^2 - z^2 . \qquad (1)$$

Interval s remains constant at usual Lorentz transformations in the (1 + 3)-D Minkowski space. However, the interval is changed at turns in the (1 + 4) extended space G (t; x, y, z, s) in planes (t; s), (x, s), (y, s) and (z, s).

In extended (1 + 4) space G (t; x, y, z, s) to each particle is compared 5-vector:

$$\overline{p} = (E/c;\ p_x ,p_y ,p_z ,mc). \qquad (2)$$

For free particles it is isotropic:

$$(E/c)^2 - (p_x)^2 - (p_y)^2 - (p_z)^2 - (mc)^2 = 0. \qquad (3)$$

Geometrical meaning of new coordinate - interval in the Minkowski space, and physically we connect it with an index of refraction n [1]. The problem of comparison to each field some refraction index in ESM is not solved yet. However, we shall mark, that within the framework of general theory of relativity a refraction index appropriated to weak gravitational field is possible to introduce too [17,18].

In ESM it is supposed, that at motion on trajectory with variable n the rest mass of particles varies, that results in change of a gravitational field, created by them.

In particular, particles with zero mass (for example, the photons) falling from empty space with n = 1 on a substance with n> 1, acquire nonzero mass and become a gravitational field source. Offered in [5,15] integrated equations system can describe such processes.

The capability that photon has nonzero mass, is widely discussed both theorists, and experimenters. The reviews of the last papers concerning this problem see in [3].

Let's mark, that the appearance of photon variable weight can be predicted and within the framework of the classical approach. For this purpose, it is necessary to consider case, when the photon spread in a resonator or waveguide [3,7].

There is an interesting prediction in frameworks ECM, which experimental check will help to evaluate correctness of interval introducing as 5-th coordinate. ESM predicts possibility of correction to departure angle of the photons of in Vavilova-Cherenkova effect. The correction is called by motion of particles along fifth coordinates s, to which there corresponds - some index of refraction n [1]. Really, at radiation of a Bremsstrahlung caused with superlight speed in environment by an electron time delay of emitting a braking gamma quantum according to ESM should happen. In the ESM frameworks it follows from that fact, that a gamma - quantum after the birth from area in the nearest neighborhood of an electron, to which corresponds interval $s_1 >> 0$ and index of refraction $n >> 1$ will move to area described by $s_0 \approx 0$ and $n \approx 1$. This will take time of order $t \approx (s_1 - s_0)/c$.

Such departure time delay of a braking gamma-quantum will cause to displacement of gamma - quantum departure point forwards in direction of electron beam motion. It will be cause to decreasing observable in experiment Cherenkov angle of gamma - quantum departure. The given prediction, certainly, requires careful experimental check. However, one from indirect reasons that it is correct prediction is the experiment on observation Vavilova-Chernkova effect in Stanford University [21].

At theoretical value of Cherenkov angle: $\varphi_{THEOR} = 2{,}4 \cdot 10^{-2}$ radian, it was obtained on 10 % smaller experimental value of Cherenkov angle: $\varphi_{EXPERIM} = 2{,}2 \cdot 10^{-2}$ radian.

Close to ESM approaches to construction (1 + 4) D spaces are model proposed in [4] Here as fifth the coordinates are offered to be used mass (substance). However, as authors recognize in this model, it is impossible to construct, for example, energy-momentum tensor. In ESM this defect is absent [5,6].

In ESM were constructed a mechanics of the mass points and electrodynamics [1] , are considered 5-D Lienard-Wiechert potentials [2]. In ESM frameworks the main gravitational effects considered in a general theory of relativity were obtained [16]. It is shown, that the ESM give the same results in first order, as general theory of relativity.

In ESM photon have mass when photon is in the area characterized by s> 0 and n>1. This fact allows us to take into consideration objects, which properties are identical to the dark matter and dark energy [15]. The problems connected to existence of dark matter and dark energy and their probable transformation to each other are widely discussed now in connection with appearance during last years large a number of new astronomical observations and results [8,9,10,14,19].

Simultaneously with development of a theoretical ESM part the experiments on checking of a qualitative prediction in ESM frameworks concerning a capability of photons transformation in fields of space with n>>1 into new objects were conducted in the [6,11].

Example of such process - hit a gamma-quantum in a nuclear nucleus and consequent processes called by this event.

For fields ESM predicts transformation electrical $\vec{E}$ and magnetic $\vec{H}$ fields in new fields: vectors field $\vec{G}$ and scalar field Q [1,6,11,13].

The first series of experiments, conducted in 2001 and described in [6,11,12,13] has given a positive result. The authentic deviation of a torsion pendulum from the fixed braking tungsten target was revealed statistically. The target was irradiated by gamma-quantum arising at breaking electrons, in the substance. The target was irradiated with an electron beam with energy 30 MeV and mean beam power equal 450 W. In this experiment, there was a force equal to $10^{-6} – 10^{-8}$ N between the stopping target and one of the loads at the end of a torsion pendulum.

In 2001 two essential features of arising effect were also detected:

-First, the arising force calls a repulsion of torsion pendulum load from the stopping target (more exactly - repulsion from the stopping target nearest of the torsion pendulum load).

-Secondly, effect of repulsion did not disappear at once after the electron beam cut off. There was rather continuous aftereffect compared in time with time of stopping target irradiation by electrons (about 10 minutes).

## Experimental set up description and scheme of measurement.

With taking into account first experiments (in 2001) new experimental set up in August - September 2004 was created.

The scheme of measurements was essentially changed Fig.1.

Two heavy loads were balanced each other. Loads were situated at the different ends of a horizontal pendant. The loads could move only in vertical plane. In the experiment, we measured force, which can arise as result of loads disbalance in the Earth gravitational field during irradiation one of loads by an electron beam.

Horizontal pendant for loads was duraluminium U-beam 1 with length 200 cm and cross section 4x3x4 cm. On the long horizontal pendant end 1 was the braking duraluminium target 2 with weight 3300 g and shape of cylinder with diameter 12 cm and length 10,5 cm.

Horizontal pendant was attached to a bronze slice 3The slice 3 was attached to horizontal pendant 1 thus that slice center with an angular cut-out on the slice bottom was in the point 4. The point 4 has divided horizontal pendant in the ratio 1:3.The slice 3 was established triangular hole downwards and can turn on bronze support 5 with triangular cross section. The apex angle of a triangular support was equaled to 60 degrees. The support 5 was attached to an aluminium slice 6.

The slice 6 was loaded on boundaries with several massive leaden freights. The freights densely pressed a slice 6 to a metal rack 7. The rack 7 was loaded with freight from leaden bricks of gross weight in 1500 kg. The rack 7 served also protection electronic balance 11 from radioactive radiation.

On the short horizontal pendant end 1 there was a second load 8 with weight equal to 11670 g. This load counterbalanced the braking target 2. The precisely balancing of loads 2 and 8 was made with the help of small mobile load 9 with weight 110 g. All elements of the experimental installation were made from unmagnetized metals and carefully grounded with the help of copper wires.

Weight of fixed load 10, located under the oscillating target 2 was 1008 kg. The distance between the loads 2 and 10 was equaled 0,5 cm. The fixed load 10 was combined from leaden bricks and had the horizontal sizes 30x40 cm and altitude 70 cm.

Load 8 lied on electronic balance cup 11. The loads 2 and 8 oscillation on the horizontal pendant ends 1 did not exceed 1 mm in a vertical direction.

With the help of mobile load 9 was installed such balance of equilibrium between loads 2 and 8, that electronic balance 11 showed no more than 8-12 g. This small unbalance between loads 2 and 8 was quite enough for a stable position of load 8 on a balance cup 11.

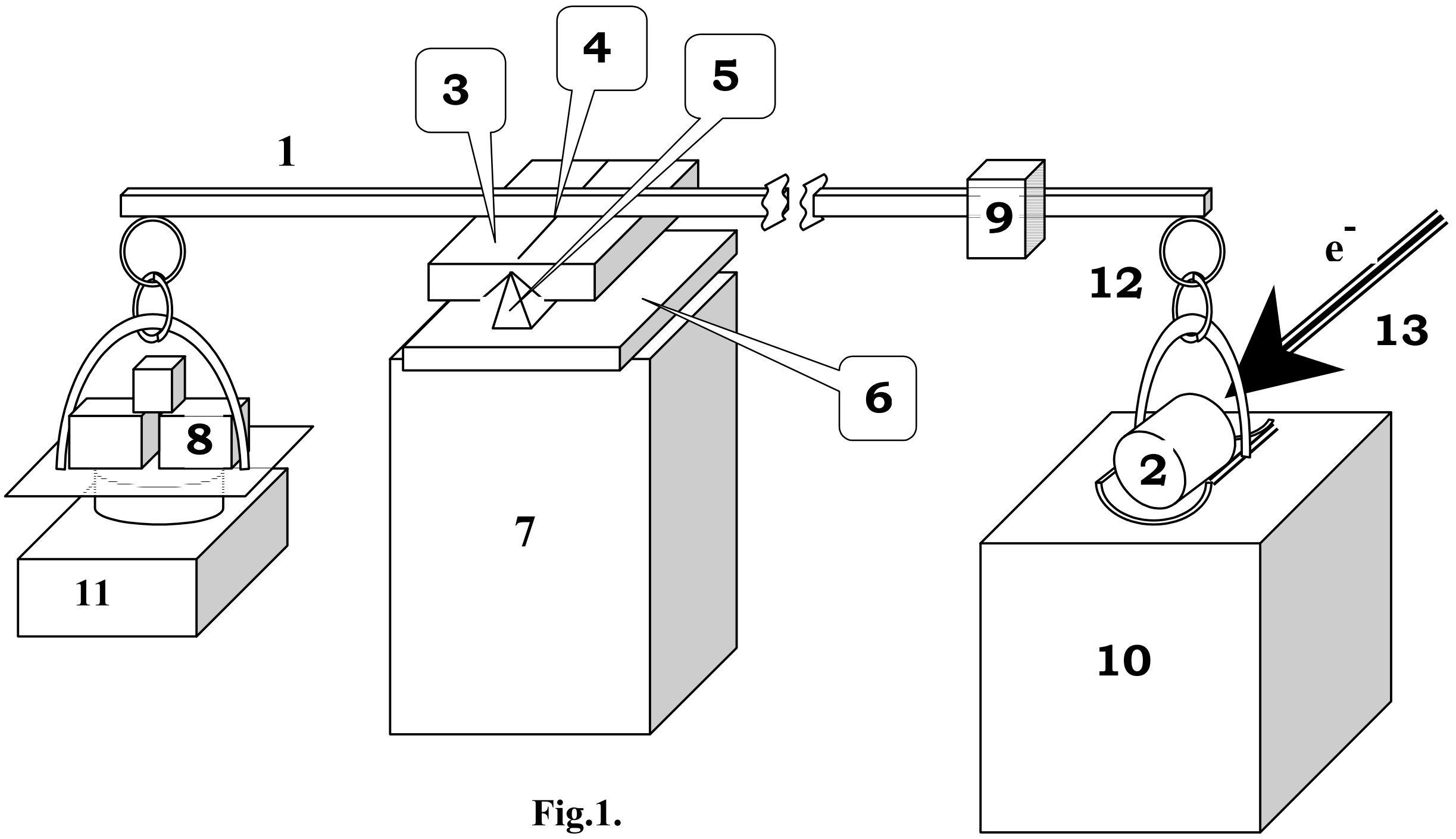

Fig.1.

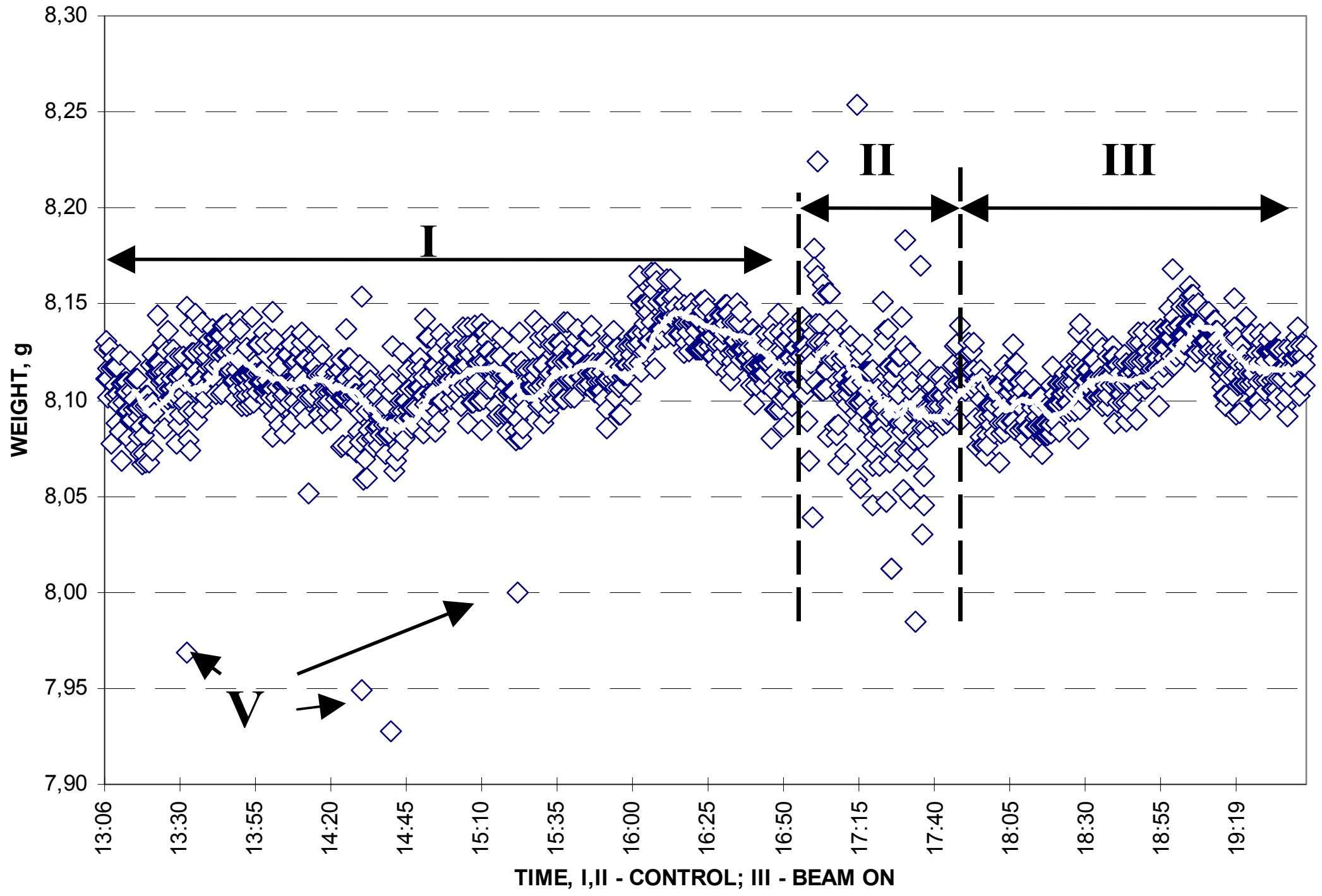
8,30
8,25
8,20
8,15
8,10
8,05
8,00
7,95
7,90
WEIGHT, g
I
II
III
V
13:06
13:30
13:55
14:20
14:45
15:10
15:35
16:00
16:25
16:50
17:15
17:40
18:05
18:30
18:55
19:19
TIME, I,II - CONTROL; III - BEAM ON

Fig.2.

The accuracy of electronic balance CAS MW-150 is 0,005 g, and the limiting weight is equal 150 g. Optimum for balance was the load in range no more than 10 % from a maximum limit of weighing.

Thus, horizontal pendant 1 with loads 2 and 8 had two rallying-points: the main load (about 15980 g) was come on a support 5 with a triangular profile and length 10 cm, and the small weight less then 0,05 % from total load was enclosed in a electronic balance cup 11. Free there was only pendant 1 long arm with braking target 2 attached to it in the bottom. The braking target 2 is stable lied on the semicircular basis. The basis was made of a thin copper sheet attached by a thick copper wire with section 0,5 cm to pendant 1. The attachment to pendant 1 was made from two mutually perpendicular aluminium well polished rings of 12 with diameter 2,5 cm.

The electron beam 13 from the accelerator had energy 50 MeV, frequency 10 Hz, pulse duration 4 ms, mean energy 10 W and diameter 15 mm. The beam 13 at first was partially braked in a thin tungsten plate with thickness of 2,5 mm. Then radiation consisting of electrons Bremsstrahlung (gamma - quantum with energy $< 50$ MeV) and electron beam which partially lost energy, hit in the end face duraluminium of the braking target 2.

Hit of an electron beam in the braking target 2 we inspected on current value going from the braking target on ground through load resistance. The irradiation was conducted in a horizontal plane in perpendicular direction to pendulum 1 oscillation. The oscillation of mean electron beam power in time was about 20 % and was due to features of the accelerator design.

## Calibration measurements

Within several weeks the careful measurements of pendulum behavior with loads in various conditions were conducted. It was made for study of all probable alternate effects on experimental results.

Such possible influence can be change of air flows at turning the power on and off of forced cooling or at closing and opening of metal doors of radiation protection. The results could be influenced by change of temperature and humidity in a premise or direct heating of the target at stopping in the target of electrons and gamma-quantum.

The influence called accumulation of static electrical charges and effect of mechanical oscillations are possible too. The mechanical effects are probable from different activities such as movement of the people in a building and machines in the street and from braking and boost on the nearest station underground trains.

All these possible spurious effects on the experimental installation were carefully investigated.

The main idea at realization of calibration measurements was that to us is difficult precisely to calculate an operation of various external effects at complex system consisting from mechanical pendulum with loads and electronic balance.
However we can beforehand measure or to simulate this effect and thus to distinguish a useful signal from changes, called by other reasons.

On Fig.2 one is one of such model experiments. On the Fig.2 the balance indications up to moment of accelerator switch on - area I from 13:00 until 17:40 are adduced. In this period in the experimental hall the starting-up and adjustment works were also conducted.
On Fig. 2 area II - from 16:50 until 17:40 activities are carried out in immediate proximity from the installation.

Area III - from 17:41 until 19:42 doors of protection are closed, the cooling is included, the accelerator works with mean beam power 10 W. However, in III all radiation is swallowed by leaden absorber 10 cm thickness, situated directly before the target 2.
Area V is single mechanical disturbances by duration no more than 15 seconds.

It is visible, that within the limits of experimental set up sensitivity any noticeable influence on measurements results the accelerator activity does not render (area III).

The signal remains in limits [8,120 + /- 0,025] g, despite of various effects on measurements results from electrical, thermal, mechanical, radiation and other handicaps during all time of observation 6 hours 40 minutes.

## Experimental results

A series of experiments was conducted 01.10.2004-14.02.2005.

On Fig.3 relative change of equilibrium between loads 2 and 8 are adduced. The results are obtained at irradiation of the target during 3 h 50 m by an electron beam with mean power 10 - 12 W. On Fig.3 the monitoring measurements before and after irradiation are adduced also.

The measured results before beam actuation from 10:05 till 12:00 (area I) and in 90 minutes after beam cutoff from 17:50 till 21:30 (area VI) lie in limits [9,125 + /- 0,03] g.

It is visible, that on a comparison with control levels before and after irradiation the signal begins to grow almost simultaneously with a moment electronic beam switching on at 12:01 and in 100-110 minutes about 13:50 signal achieved maximum level equal to [9,34 + /- 0,025] g - area II.

Further signal remains almost 90 minutes constant till 15:25 - area III. After that moment decreasing mean power of an electron beam from the accelerator approximately on 20% was fixed.
In this moment of the electronic balance indications smoothly began to decrease within 35 minutes from 15:26 till 16:00 (area IV) up to a moment electron beam cutoff at 16:01. After cutoff of the electron beam from 16:02 till 17:50 (area V) the signal has become to decrease further and about 17:50 signal level practically has coincided with a level prior to the beginning irradiation. After 17:50 signals already was not changed and remained in limits [9,135 + /- 0,025] g. On Fig.3 area VII is a disturbance from starting-up and adjustment works in an experimental hall.

The absolute change of the signal was equal to [0,20 + /- 0,04] g. With allowance for ratios of pendulum arms 1:3 these changes are equivalent to appearance of additional force, equal [6,7 + /- 1,3] $*10^{-5}$ N. This additional force is directed vertically up to the braking target 2.

On Fig.4 the results of relative equilibrium change between loads 2 and 8 are adduced. These results are obtained at target 2 irradiation by an electron beam with mean power of 4-5 W within 2 hours. On the Fig. 4 control measurements before and after irradiation are adduced also.

Control measurements before (area I - from 16:07 up to 18:01) and after (area IV - from 21:21 up to 24:00) irradiations lie in range [11,10 + /- 0,025] g.

Area II - from 18:02 until 20:02 measurements results, obtained during target 2 irradiation. Area III - from 20:02 until 21:20 aftereffect and gradual recovery of signal control level.

On Fig.4 the disturbances from starting-up and adjustment works conductive in an experimental hall before and after irradiation of the target are visible also – area V.

The value of disbalance signal in 110 mines after a beginning of irradiation has achieved [11,20 + /- 0,025] g.

Fig.3.

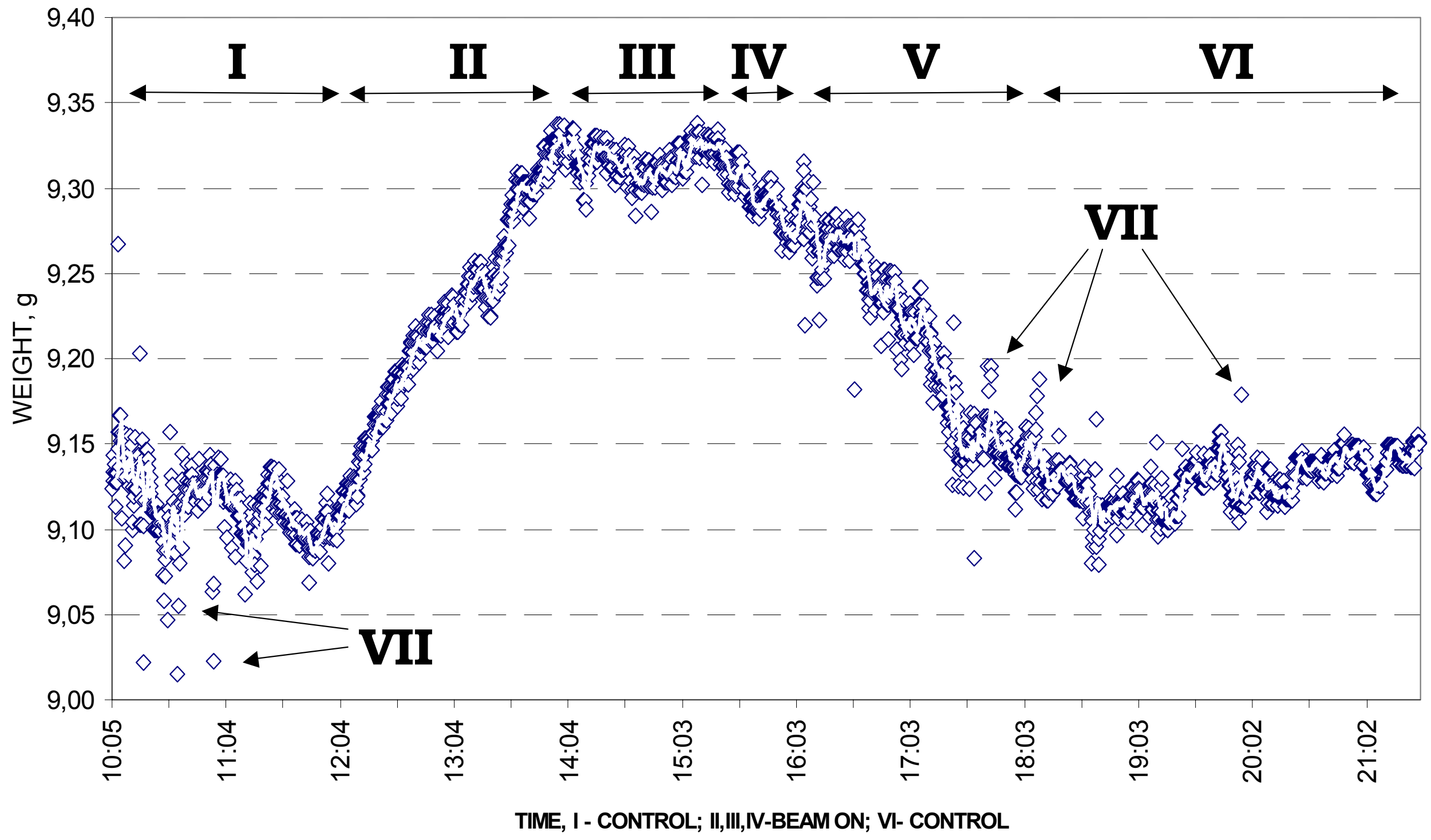
I
II
III
IV
V
VI
VII
VII
9,40
9,35
9,30
9,25
9,20
9,15
9,10
9,05
9,00
WEIGHT, g
10:05
11:04
12:04
13:04
14:04
15:03
16:03
17:03
18:03
19:03
20:02
21:02
TIME, I - CONTROL; II,III,IV-BEAM ON; VI- CONTROL

Fig.4.

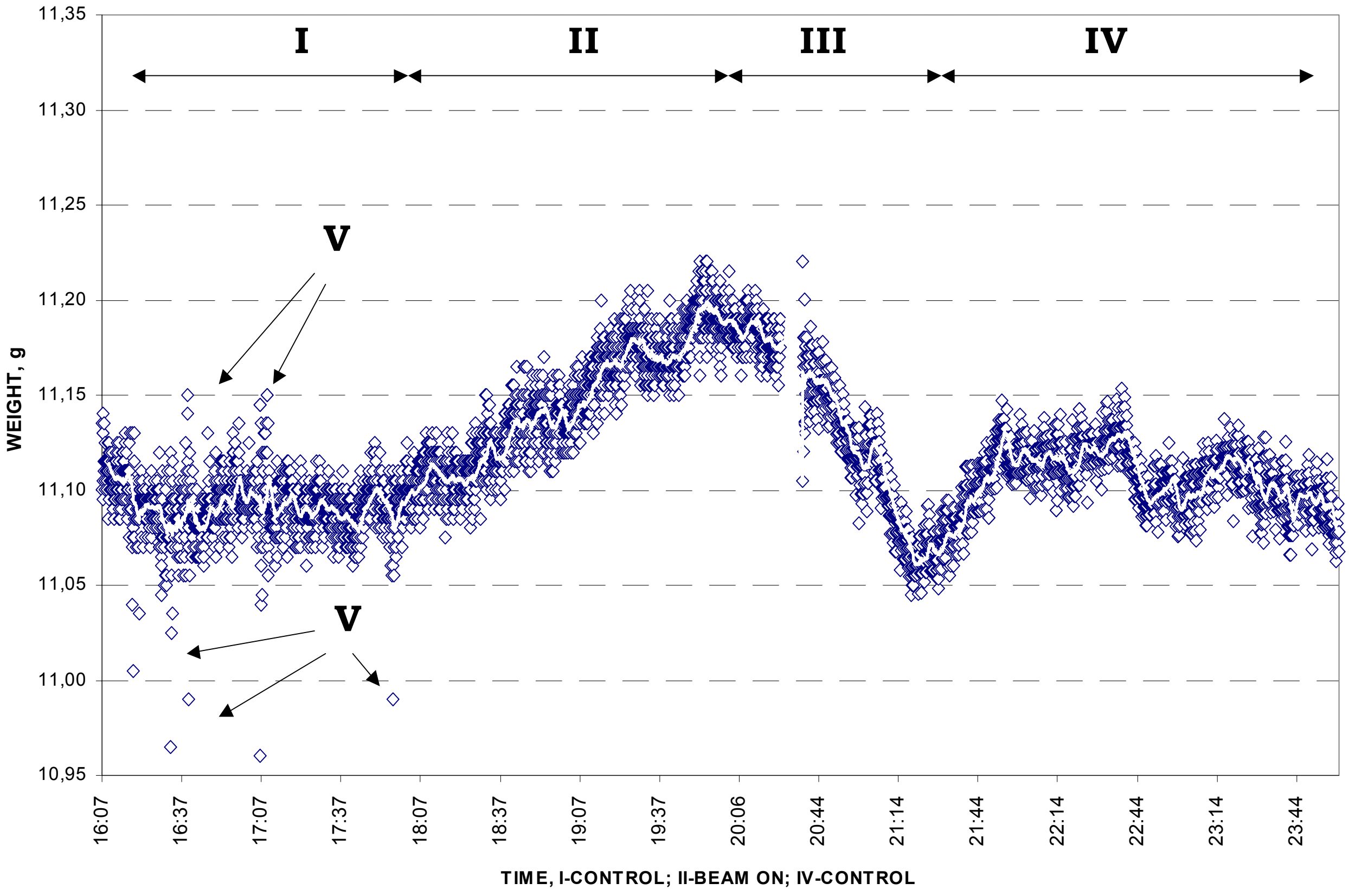

The absolute change of disbalance signal was [0,10 + /-0,035]. Has a place also aftereffect within approximately 1 hour 20 minutes after electron beam cutoff.

## Discussion

To interpret obtained results within the framework of extended space model or any other approach, we should at first be convinced, that the force, arising at irradiation, is not connected with some other reasons.

The effect of accumulation of electrostatic charges in pendulum 1, target 2 or leaden loads 10 under it was investigated. We found insignificant influence of directed electrostatic charges to measurements results.

Influence to the electronic balance indications of air flows which can arise in an experimental hall was investigated also. At opening/ closing of protection doors no any noticeable changes, in the balance indications it is not revealed.

The probable direct heating of the target of the installation by activity of the accelerator also can call change of the balance indications. However it was revealed, that the power of the heater at 5-10 of time large than mean power of an electron beam is necessary for obtaining noticeable effect.

As to a capability of influence on the balance indications of various mechanical oscillations transmitted by the ground from machines, underground or other working mechanical devices, this effect not essentially taking into account:

- at first, coincidence beginning time of indications changes with the moment of electron beam switch on,
- secondly, various measured duration and relation of an absolute value of effect to duration of irradiation.

Any correlation of the balance indications with street weather conditions, such as: temperature of an air, pressure, humidity and speed of a wind was not revealed.
For this purpose the signal from balance continuously recorded during two series by duration for 7 day each and indications of meteorological station that was most close to the place of experiments realization was compared.

Qualitatively results of experimental series in 2004 and 2001 were coincided. In both experimental series, the origin of a repulsive force during irradiation is revealed and the effect of an aftereffect comparable on duration with total time of irradiations was detected.

With allowance for that in 2001 we had a horizontal torsion pendulum, and in 2004 experiments were conducted in a vertical plane with practically fixed loads, the contingency of results coincidence is represented improbable.

Besides, experiments 2004 and 2001 were conducted in completely different places. The accelerators had different parameters of electron beams. Thus no possible spurious external effect, not taken into account in calibration measurements, could be identical.

It is important to notice, that the obtained effect is proportional to time of irradiation (absorbed braking target doze). It follows from comparison results of two target irradiation series during 3 hours 50 min (Fig.3) and 2 hours (Fig.4).

## Conclusions

Repeated series of experiments 2004 on check of prediction about possible transformation fields, made within the framework of extended space model is conducted. A repeated series of experiments in 2004 has given results qualitatively coincides to results of the first experimental series in 2001.

Has a place effect of repulsion between loads 2 and 10, at irradiation one of the loads by a Bremsstrahlung of electrons with energy 50 MeV.
The reached effect is called by origin of repulsive force of order [6,7 + /- 1,3] $*10^{-5}$ N. The value of the effect is proportional to radiation doze absorbed braking by the stopping target. In 80 - 110 minutes after the beginning of irradiation has the place effect of saturation. The effect of aftereffect comparable on time with irradiation time of the braking target by electron beam is detected also.

The original scheme of experiment has allowed in real-time mode to weigh the braking target immediately during irradiation.

The applied scheme of experiment has enabled maximum to simplify geometry of measurements. In the experiments is used almost fixed pendulum with one degree of freedom varying in a vertical plane (2004) instead of torsion pendulum having by 6 degree of freedoms (2001).

Undoubtedly clear interpretation of obtained results within the framework extended space model will require realization of additional experiments and will be a subject of consequent activities.